\documentstyle[12pt]{article}
\advance\hoffset by -1.1truecm
\advance\voffset by -1.0truecm

\title{String/Quantum Gravity motivated Uncertainty Relations and 
Regularisation in Field Theory\footnote{To appear in Proc.
XXI Int. Coll. on Group Theor. Methods in Physics, 
Goslar, July 1996 }}
    \author{Achim Kempf\thanks{Research Fellow of 
Corpus Christi College in the University of Cambridge, $\mbox{
\qquad \qquad \qquad ~\quad}$ $\mbox{~~~~}$ 
Email: a.kempf@amtp.cam.ac.uk} \\
Department of Applied Mathematics \& Theoretical Physics\\
University of Cambridge\\ Cambridge CB3 9EW, U.K}
\date{}
\def\be{\begin{equation}}
\def\ee{\end{equation}}
\def\ba{\begin{eqnarray}}
\def\ea{\end{eqnarray}}

\newcommand{\x}{{\bf x}}
\newcommand{\p}{{\bf p}}

\newcommand{\sn}{\smallskip\newline}

\newcommand{\bn}{\bigskip\newline}
\newcommand{\mbo}{{\mbox{ }}}

\begin{document}
\maketitle
\vskip-8truecm

\hskip11.5truecm
{\tt DAMTP/96-101} 

\hskip11.5truecm
{\tt hep-th/9612082}
\vskip7.4truecm
\section{Overview} 
The uncertainty relations and the underlying
canonical commutation relations are at the heart of quantum mechanics. 
In recent years, for various conceptual and technical reasons,
there has been renewed interest in the
question whether there may exist corrections to the canonical
commutation relations which could be significant at extreme scales. 
Let me start with a brief (and certainly incomplete) overview.
\sn                             
Probably the most general approach 
is the ansatz of `generalised quantum dynamics', 
developed by Adler et al. 
This framework not only allows for commutation relations of generic form,
but also includes a possible generalisation of the normally
underlying complex Hilbert space to a quaternionic or
octonic space. Within this approach,
the ordinary canonical commutation relations have been shown to arise 
as a first order approximation from a statistical 
averaging process, see \cite{adleretal}.
\sn
Studies which suggest specific correction 
terms to the commutation relations between the position and the 
momentum operators have appeared in the context of 
both string theory and quantum gravity.
From the quantum gravity point of view it has long been argued that,
when attempting the resolution of extremely small distances,
the space-time disturbing gravitational effect of the 
necessarily high energy of the probing particle
must eventually pose an 
ultimate limit to the possible resolution of distances, the latest
at the Planck scale. Indeed, in string theory, a number of studies, 
e.g. on string scattering, have suggested the existence of a 
finite minimal uncertainty in positions $\Delta x_0$, see e.g.
\cite{grossmende,amati,maggiore}. Intuitively, 
the use of higher energies for probing small scales eventually 
no longer allows to further improve the spatial 
resolution since this energy would enlarge the probed string.
The net effect has been found to be a correction to the $\x,\p$ uncertainty
relation of the form
\be
\Delta x \Delta p \ge \frac{\hbar}{2} (1 + \beta (\Delta p)^2 +...)
\label{eucr}
\ee
Through  
$\Delta A \Delta B \ge 1/2 \vert\langle [ A,B]\rangle\vert$ 
then follows $[\x,\p] = i \hbar (1 + \beta \p^2 +...)$.
As is easily checked, Eq.\ref{eucr} implies 
a finite minimal uncertainty $\Delta x_0 = \hbar \sqrt{\beta}$.
For recent reviews see \cite{witten,garay}. 
The existence of a lower bound $\Delta x_0$ to the standard deviation of 
position measurements would be a
true quantum structure, 
in the sense that it has no classical analog. Among its
attractive features is that it does not require the breaking of translation 
invariance. An additional 
scale dependence of $\Delta x_0$, 
due to time-of-flight effects, has also been suggested, for a recent
reference see e.g.\cite{gac}.
\sn 
Specific correction 
terms to the commutation relations among the position operators
have also been suggested, in particular, in \cite{dfr}. 
A key idea there is, that for optimal spatial resolution in one direction,
the energy of the probing particle should be delocalised in orthogonal
directions, in order to reduce the gravitationally disturbing
energy density at the location of the measurement. Uncertainty relations
of the form $\sum_{i>j} \Delta x_i \Delta x_j \ge l_{pl}^2+...$ have 
therefore been suggested. Noncommuting position operators
were probably first investigated by Snyder in 1947 \cite{snyder},
in an approach which has been followed since, mainly by Russian schools,
see e.g.\cite{kadyshevskii}.
\sn
Specific correction terms to the commutation relations among the
momentum operators have also been suggested, e.g. in \cite{banach},
with the underlying idea to account for the
noncommutativity of translation operators on curved space.
These corrections would only be relevant at large scales, i.e. as
an infrared effect.
\sn
A related field, in the sense that it also involves generalised commutators,
is the programme of exploring the possibilities of
generalised internal and external symmetries.
This field is being pursued intensively in the literature, in the context of
noncommutative geometry in the sense of Connes \cite{connes},
and in the context of quantum groups, to which 
the special volume of these proceedings is devoted.
\section{Feynman rules on generalised geometries}
We consider general canonical commutation relations
\be
[\x_i,\p_j] = i\hbar \left(\delta_{ij} + \Theta_{ij}(\x,\p)\right)
\label{gcr}
\ee
where $\Theta$ is a not necessarily symmetric function in the $\x_i$
and $\p_i$. Similarily, we allow 
$[\x_i,\x_j] \ne 0$ and $[\p_i,\p_j] \ne 0$. 
We require however the standard involution
$\x^\dagger = \x,~ \p^\dagger = \p$ in order to guarantee real expectation
values.
\sn
In this way we cover the case of the corrections to the
$\x,\p$ commutation relations which we mentioned above, namely which
imply a finite minimal uncertainty $\Delta x_0$, as well
as the case $[\x,\p] = i\hbar (1 + \alpha \x^2 +\beta \p^2)$
which implies minimal uncertainties $\Delta x_0$ and $\Delta p_0$ 
in both positions and momenta, see \cite{ak-jmp-ucr}-\cite{ak-hh-1}.
\sn
A general framework for the formulation of quantum field theories on
such `noncommutative geometries' 
has been given in \cite{ft,ak-prd-2}.
Consider the example of euclidean charged scalar $\phi^4$-theory:
\be
Z[J] := N \int D\phi\mbo e^{-\int d^4x \mbo [
\phi^* (-\partial_i \partial^i + m^2 c^2)\phi 
 + \frac{\lambda}{4!}(\phi \phi)^*\phi \phi - \phi^*J - J^*\phi] }
\label{fto}
\ee
For our purpose it is useful to formulate the action functional 
without reference to any particular choice of basis in the space 
of fields that is formally being summed over.
The functional analytic structure is analogous to the 
situation in quantum mechanics, with fields being vectors
in a representation of the canonical commutation relations. 
We therefore formally extend the Dirac notation for states to fields, i.e. 
$\phi(x) = \langle x\vert \phi\rangle$.
Of course, the simple quantum mechanical 
interpretation of fields $\vert \phi\rangle$
and in particular of
the position and momentum operators does not simply
extend.
However, this notation clarifies the functional analytic
structure of the action functional:
\be
Z[J] = N \int D\phi \mbo e^{ - \frac{l^2}{\hbar^2} \mbo \langle 
\phi \vert  
\mbo\p^2 + m^2 c^2\mbo\vert \phi \rangle \mbo
- \frac{\lambda l^4}{4!} 
\langle \phi * \phi \vert \phi * \phi \rangle \mbo
+ \langle \phi \vert J\rangle + \langle J\vert \phi\rangle }
\label{dn}
\ee
(We introduced a unit of length $l$ which 
could be reabsorbed in a trivial field redefinition).
Ordinarily, when formulating a field theory in position space,
as in Eq.\ref{fto}, 
the fact that $\p^2$ is represented as $- \hbar^2 \nabla$ already 
implies that $\p$ is represented as $-i\hbar \partial_{x_i}$, so that
it must obey the ordinary commutation relations.
The advantage of the representation independent formulation
in Eq.\ref{dn} is that the underlying commutation relations are not 
implicitly specified and can therefore be generalised.
\sn
We can then derive the Feynman rules in any arbitrary
Hilbert basis $\{\vert n\rangle\}_n$ in the space $F$ of fields
on which the generalised commutation relations 
Eqs.\ref{gcr} are represented.
While this basis can be chosen
continuous, discrete, or generally a mixture of both,
we here use the convenient notation for discrete $n$. 
Fields, operators and the pointwise multiplication $*$ of fields
are now expanded as 
\be \phi_n = \langle n\vert \phi \rangle \qquad 
\mbox{and }\qquad 
(\p^2 +m^2c^2)_{nm} = \langle n \vert \p^2 + m^2c^2 \vert m\rangle
\ee
\be
* = \sum_{n_i} L_{n_1,n_2,n_3} \vert n_1\rangle \otimes
\langle n_2\vert \otimes \langle n_3 \vert
\ee
We remark that, ordinarily, $(\phi_1*\phi_2)(x) = \phi_1(x)\phi_2(x)$, 
i.e. $*$ takes the form:
\be 
* =  
\int d^4x ~\vert x\rangle \otimes \langle x\vert \otimes \langle x \vert
\label{el}
\ee
In the $\{\vert n\rangle\}$ basis, Eq.\ref{dn}
reads, summing over repeated indices:
\be
Z[J] = N \int_F D \phi \mbo e^{ -\frac{l^2}{\hbar^2}\mbo
\phi_{n_1}^* (\p^2 +m^2c^2)_{n_1n_2} \phi_{n_2}
-\frac{\lambda l^4}{4!} L^*_{n_1n_2n_3}
L_{n_1n_4n_5} \phi^*_{n_2} \phi^*_{n_3} \phi_{n_4}\phi_{n_5}
+ \phi^*_n J_n + J^*_n \phi_n }
\ee
Pulling the interaction term in front of the path integral,
completing the squares, and carrying out the gaussian integrals
yields 
\be
Z[J] = N' e^{-\frac{\lambda l^4}{4!} L^*_{n_1n_2n_3} L_{n_1n_4n_5}
\frac{\partial}{\partial J_{n_2}}
\frac{\partial}{\partial J_{n_3}}
\frac{\partial}{\partial J^*_{n_4}}
\frac{\partial}{\partial J^*_{n_5}}}
\mbo
e^{-\frac{\hbar^2}{l^2} J^*_n (\p^2 +m^2c^2)^{-1}_{nm} J_m}
\label{al1}
\ee
The Feynman rules therefore read, see \cite{ft,ak-prd-2}:
\be
\Delta_{nm} = \left(\frac{\hbar^2/l^2}{\p^2+m^2c^2}\right)_{nm}, \qquad
\Gamma_{rstu} = - \frac{\lambda l^4}{4!} L^*_{nrs}L_{ntu}
\label{al2}
\ee
On ordinary geometry, i.e. with the ordinary 
commutation relations $[\x_i,\p_j]=i\hbar \delta_{ij}$ underlying,
the choice e.g. of the position eigenbasis 
$\vert n\rangle = \vert x \rangle$
or the momentum eigenbasis $\vert n\rangle = \vert p \rangle$
of course recovers the usual formulations of the Feynman rules.
\section{Regularisation}
Eqs.\ref{al2} yield the Feynman rules for generic commutation relations,
and for arbitrary choices of basis. Let us now consider the question 
whether the theory is UV and/or IR finite 
on geometries (commutation relations) which imply
a finite minimal uncertainty $\Delta x_0>0$ and/or $\Delta p_0>0$.
Indeed, the following two statements can be made for arbitrary geometries
(generalised commutation relations):
\bn
{\bf (A)} $\Delta p_0>0 ~~~ \Rightarrow~~~ $ in QM: ~
$\vert\vert \frac{1}{\p^2}\vert\vert < \infty~~~ \Rightarrow~~~ $
in QFT: propagator IR regular
\bn
Both steps, first that $\Delta p_0>0$ implies that the inverse of the 
operator $\sum_i\p_i^2$ is a bounded self-adjoint operator, 
and secondly 
the implementation into field theory have been shown in \cite{ak-prd-2}.
A crucial fact is that $\p^2$ becomes positive {\it definite} on
any dense domain on which the commutation relations are represented,
if $\Delta p_0 >0$.
\bn
{\bf (B)} $ \Delta x_0>0 ~~~ \Rightarrow~~~ $ in QM: ~
$\vert\vert~\vert x^{ml}\rangle \vert\vert < \infty~~~ \Rightarrow~~~ $
in QFT: vertices UV regular
\bn
So far, in all known geometries with $\Delta x_0>0$
the states of maximal localisation
have been found to be normalisable, see 
\cite{ak-jmp-ucr,ak-gm-rm-prd,ak-hh-1}.
To see that this is generally true, assume $\vert \psi_n\rangle$
to be a sequence of physical states (i.e. they are in the domain of 
$\x$ and $\p$) which approximates the vector
of maximal localisation, say around the origin:
$\lim_{n\rightarrow \infty} \Delta x_{\vert\psi_n\rangle} = \Delta x_0$.
It is a sequence within the Hilbert space with respect to the norm
induced by the now positive {\it definite}
 $\x^2$, and converges therefore 
towards a vector
$\vert \psi\rangle $ within that Hilbert space. Since
the $\x^2$-induced norm is sharper than the original Hilbert space norm,
$\vert\psi\rangle$ is normalisable. Preliminary results on 
UV regularisation through $\Delta x_0>0$ are in \cite{ft}.
I am currently working out a more comprehensive study 
with my collaborator G. Mangano \cite{ak-gm-2}. 
Here is a sketch of the main points.
\sn
In Eq.\ref{dn} the pointwise multiplication $*$ of fields 
is crucial for the description of local interaction and
is normally given through Eq.\ref{el}, yielding
$ (\phi_1 * \phi_2)(x) = \phi_1(x) \phi_2(x)$.
On generalised geometries, in order to describe maximally local 
interactions, Eq.\ref{el} can be read with the $\vert x\rangle$
denoting the vectors of then maximal localisation, i.e.
\be 
* = 
\int d^4x~ \vert x^{ml}\rangle \otimes \langle x^{ml}\vert 
\otimes \langle x^{ml} \vert
\label{li}
\ee
where $\vert x^{ml}\rangle$ denote the fields which are maximally 
localised with position expectation values $x$,
yielding the structure constants
$L_{n_1,n_2,n_3} = \int d^4x ~ \langle n_1\vert x^{ml}\rangle
\langle x^{ml}\vert n_2\rangle \langle x^{ml}\vert n_3 \rangle$.
As abstract operators, i.e.
without specifying a Hilbert basis in the space of fields 
the free propagator and the lowest order vertex then read,
choosing $l := \Delta x_0$:
\be
\Delta = \frac{\hbar^2}{(\Delta x_0)^2 ({\bf p}^2+m^2c^2)}
\label{fr1}
\ee
\begin{equation}
\Gamma = - \frac{\lambda}{4!} \int 
\frac{d^4x~d^4y}{(\Delta x_0)^{8}}~~
\langle y^{ml}\vert x^{ml}\rangle~~
\vert y^{ml}\rangle \otimes \vert y^{ml}\rangle
\otimes
\langle x^{ml}\vert\otimes\langle x^{ml}\vert
\label{fr2}
\ee
The crucial observation is that in combining the Feynman rules to form
graphs, all factors that can appear are either of the type 
$\langle x^{ml}\vert y^{ml} \rangle$ or 
$\langle x^{ml}\vert (\p^2+m^2c^2)^{-1}\vert y^{ml} \rangle$. 
Both factors are now well behaved and {\it bounded} 
functions of $x$ and $y$, because $1/(\p^2+m^2)$ is a bounded
self-adjoint operator and, crucially, because the $\vert x^{ml}\rangle$
are normalised. Therefore, all graphs are ultraviolet regular.
Of course, as $\Delta x_0 \rightarrow 0$, both factors
converge towards the distributions as which they are normally
defined.

\end{document}